\def\year{2022}\relax
\documentclass[letterpaper]{article} % DO NOT CHANGE THIS
\usepackage{aaai22}  % DO NOT CHANGE THIS
\usepackage{times}  % DO NOT CHANGE THIS
\usepackage{helvet}  % DO NOT CHANGE THIS
\usepackage{courier}  % DO NOT CHANGE THIS
\usepackage[hyphens]{url}  % DO NOT CHANGE THIS
\usepackage{graphicx} % DO NOT CHANGE THIS
\usepackage{natbib}  % DO NOT CHANGE THIS AND DO NOT ADD ANY OPTIONS TO IT
\usepackage{caption} % DO NOT CHANGE THIS AND DO NOT ADD ANY OPTIONS TO IT
\DeclareCaptionStyle{ruled}{labelfont=normalfont,labelsep=colon,strut=off} % DO NOT CHANGE THIS
\usepackage{algorithm}
\usepackage{algorithmic}

\usepackage{amsmath}
\usepackage{amssymb}% http://ctan.org/pkg/amssymb
\usepackage{pifont}% http://ctan.org/pkg/pifont
\newcommand{\xmark}{\ding{55}}%
\usepackage{newfloat}
\usepackage{listings}
\usepackage{xspace}
\floatstyle{ruled}
\newfloat{listing}{tb}{lst}{}
\floatname{listing}{Listing}

\title{Fusion with Hierarchical Graphs for Mulitmodal Emotion Recognition}

\author{
    Shuyun Tang$^{1}$,
    Zhaojie Luo$^2$$^\dagger$$^\ast$, 
    Guoshun Nan$^3$$^\ast$, 
    Yuichiro Yoshikawa$^2$,  
    Ishiguro Hiroshi$^2$  
}

\affiliations {
    \textsuperscript{\rm 1} UC Berkeley
    \textsuperscript{\rm 2} Osaka University
    \textsuperscript{\rm 3} Singapore University of Technology and Design\\
    shuyuntang@berkeley.edu, luo@irl.sys.es.osaka-u.ac.jp, nanguoshun@gmail.com,
    yoshikawa@irl.sys.es.osaka-u.ac.jp,
    ishiguro@irl.sys.es.osaka-u.ac.jp
    
}

\usepackage{bibentry}
\begin{document}

% \maketitle

\maketitle

\begin{abstract}
Automatic emotion recognition (AER) based on enriched multimodal inputs, including text, speech, and visual clues, is crucial in the development of emotionally intelligent machines. Although complex modality relationships have been proven effective for AER, they are still largely underexplored because previous works predominantly relied on various fusion mechanisms with simply concatenated features to learn multimodal representations for emotion classification. This paper proposes a novel hierarchical fusion graph convolutional network (HFGCN) model that learns more informative multimodal representations by considering the modality dependencies during the feature fusion procedure. Specifically, the proposed model fuses multimodality inputs using a two-stage graph construction approach and encodes the modality dependencies into the conversation representation. We verified the interpretable  capabilities of the proposed method by projecting the emotional states to a 2D valence-arousal (VA) subspace. Extensive experiments showed the effectiveness of our proposed model for more accurate AER, which yielded state-of-the-art results on two public datasets, IEMOCAP and MELD.
\end{abstract}

\let\thefootnote\relax\footnote{$^{\dagger \star}$ Dr. Zhaojie Luo is the co-first and corresponding author.}
\let\thefootnote\relax\footnote{$^{\star}$ Dr. Guoshun Nan is the first corresponding author.}

\section{Introduction}
Emotions color a language and is a necessary ingredient for human-to-human communications. The objective of automatic emotion recognition (AER) is to detect emotions from enriched input signals, such as audio, texture, facial images, or multimedia signals, and it is critical for building emotionally intelligent machines. Recent years have witnessed a growing trend for the application of AER systems, such as assisting conversation \cite{morrison2007ensemble}, detecting extreme emotions \cite{kuppens2012emotional}, and  lifelike human--computer interactions \cite{cowie2001emotion,fragopanagos2005emotion}.

AER has been extensively studied for audio \cite{xu2021speech}, text \cite{calefato2017emotxt}, facial clues \cite{chen2016multithreading,luo2017facial}, and EEG-based brain waves \cite{tripathi2017using}. Previous studies showed that humans rely more on multiple modalities than on a single modality \cite{shimojo2001sensory} to better understand emotions. As highlighted in \cite{sebe2005multimodal}, voice calls are more informative than text messages, indicating that the affective prosody of audio is able to deliver additional information for emotion recognition. Meanwhile, speaking face-to-face is more effective than voice calls \cite{drolet2000rapport}, indicating that visual cues may contribute more to emotion classification.

\begin{figure}[t]
\centering
\includegraphics[width=1\columnwidth]{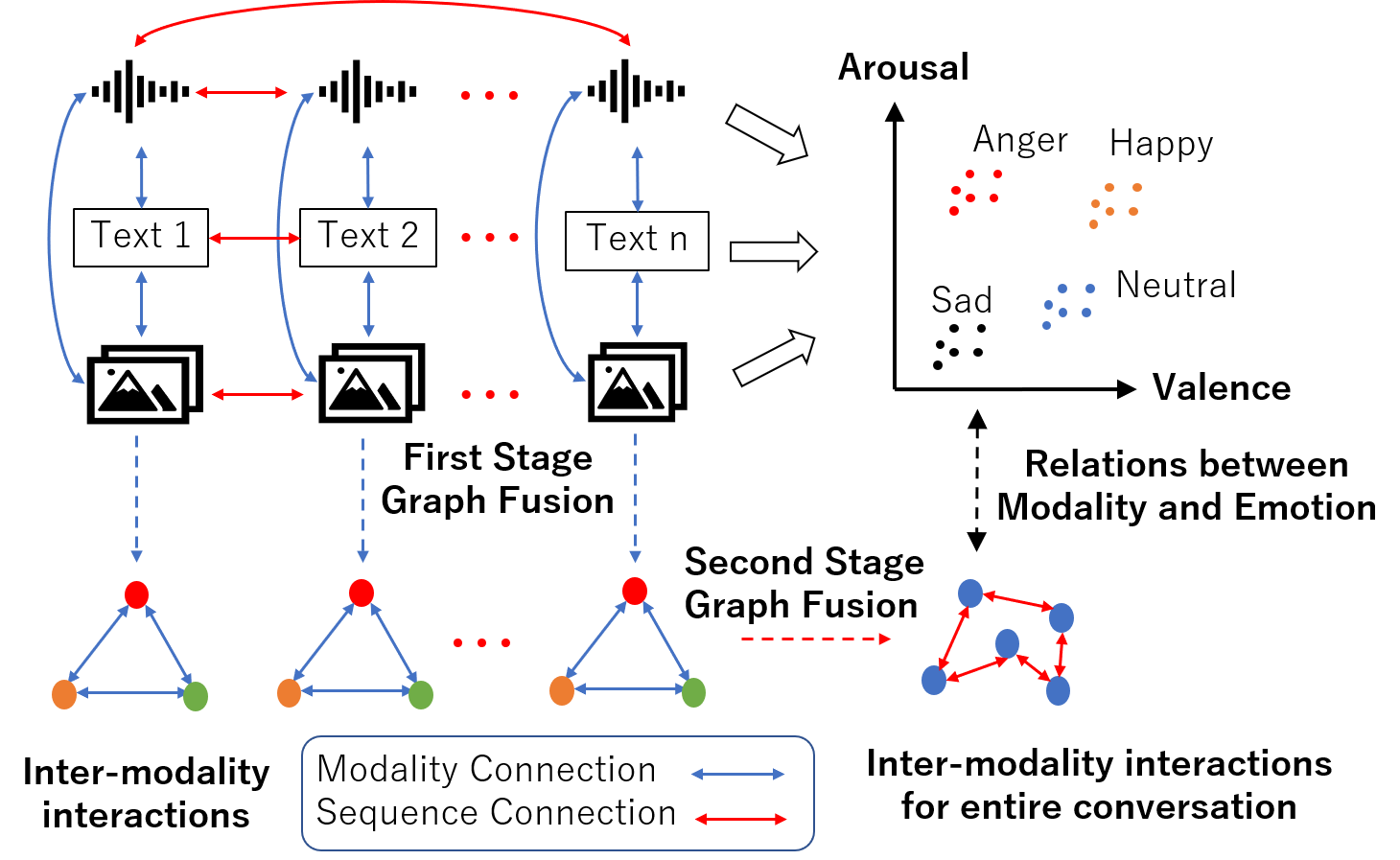} 
\vspace{-8pt}
\caption{Basic overview of our approach to multimodal emotion recognition.}
\label{fig1}
\vspace{-13pt}
\end{figure}

However, the question of how to effectively associate different modalities to the AER task is still an open research problem. The underlying reason is that different modalities are neither completely independent nor correlated, posing challenges for selecting useful information and filtering out the noise in different modalities during fusion. Previous fusion methods towards this direction used various fusion mechanisms with simply concatenated features to learn multimodal representations. These approaches can be categorized into late fusion \cite{poria2017context, xue2017bayesian}, early fusion \cite{sebastian2019fusion}, and hybrid fusion \cite{pan2020multi}.

Despite the effectiveness of the above fusion approaches, the interactions between modalities (\textbf{intermodality interactions}), which have been proved effective for the AER task \cite{fu2020multi}, are still largely underexplored. The intuition behind this is that visual, audio, and textual information are produced by a unique event, and therefore there are strong relationships between each subset of the three modalities. Recently, a graph fusion method, MMGCN \cite{hu2021mmgcn}, was proposed for modeling the intermodality interactions for emotion recognition in conversations. However, in that work, the \textbf{relations between modality and emotion} were not utilized. In reality, people tend to have more facial expressions when excited while expressing frustration through conversations. We assume that the contribution of each modality to the emotion classification depends on the types of emotions, and the relations between modality and emotion also significantly influence emotion recognition. With this in mind, our goal is to capture  both the interactions among modalities and the relations between modality and emotion for improving the AER accuracy. 

To achieve the abovementioned goal, in this paper, we propose a Hierarchical Fusion Graph Convolutional Network (HFGCN) consisting of a two-stage graph construction and a multitask learning loss based on Valence-Arousal (VA) degrees. Fig.~\ref{fig1} provides a basic overview of the proposed approach with multimodal inputs. 1) In the first stage, each modality is initialized as a node connected to another via attention-based edges, which \textbf{encode intermodality interactions into utterance representations}. Compared with previous works that mainly rely on concatenation-based fusion mechanisms, our approach leverages the inherent advantages of graphs to model the relationships between different modalities and distinguish their contribution to each utterance representation. 2) In the second stage, each utterance representation or the first-stage graph is connected to another set of attention-based edges that \textbf{extend the intermodality interactions from each utterance to the entire conversation}. 
In addition, the designed edge relations are assigned to all the edge sets for relational graph transformation, which \textbf{captures the relations between modality and emotion}.

Furthermore, we also revisit the VA theorem \cite{Barrett1998-BARDEO,barrett2006solving} and simultaneously predict the emotional states and VA degrees with a multitask learning loss,  with the aim of modeling the interactions between different types of emotions and their transitions. We empirically observe that introducing the VA degrees to map the emotional states can also facilitate the model interpretation. Extensive experiments on two public datasets show the effectiveness of our proposed method \footnote{Our code will be released upon acceptance of the paper.}. 

Our contributions are threefold:
\begin{itemize}
\item We propose a novel graph fusion network called HFGCN that is able to hierarchically fuse multimodal information by encoding the intermodality interactions into the emotion representations.
\item To the best of our knowledge, this is the first attempt to incorporate the VA degrees to AER in a multitask fashion. Such an approach can better explore the relations between modality and emotion and enhances the emotional states' interpretability. 
\item We conduct extensive experiments on two public multimodal emotion recognition datasets: IEMOCAP \cite{busso2008iemocap} and MELD \cite{poria-etal-2019-meld}.  Our experimental results show the superiority of our proposed method against previous state-of-the-art methods.
\end{itemize}

\section{Related Work}

\subsection{Multimodal Fusion}
Most multimodal fusion methods are based on the following: 
(i) Early fusion: The features of different modalities are extracted and concatenated to obtain a feature vector, which is fed into a classifier to obtain the emotion prediction. \cite{sebastian2019fusion} utilized early fusion and fed the extracted features through a convolutional neural network. 
(ii) Late fusion: The features of different modalities are extracted and directly fed into independent unimodal classifiers. Then, the intermediate predictions are concatenated and fed to a metaclassifier for the final prediction \cite{poria2017context}.  
(iii) Hybrid fusion: Combining (i) and (ii) with more than one network at different levels. \cite{majumder2018multimodal} utilized this method to first fuse each of the two modalities, and then fused the intermediate outputs and passed them to the final network. The key differences between our proposed method and conventional methods are as follows: 1) our novel fusion method is able to embed the intermodality interactions, which have proven effective for AER. Capturing such interactions with existing concatenation-based fusion methods may not yield satisfactory performance for the task; 2) our two-stage graph construction extends the intermodality interactions from each utterance to the entire conversation.

\subsection{Graph Neural Network}
Graph convolutional networks (GCNs) \cite{kipf2016semi}, which have been successfully used to address various problems in computer vision \cite{Shen_2018_ECCV} and natural language processing \cite{yao2019graph, nan2020reasoning}, provide us with a new solution to address the emotion recognition problem. Several studies have applied the GCN to speech emotion recognition works \cite{shirian2020compact, ghosal2019dialoguegcn}. However, they are only useful for unimodal tasks. Some recent works \cite{hu2021mmgcn,wei2019mmgcn} use the graph fusion method for multimodal emotion recognition without exploring the relations between modality and emotion. In contrast to the above works, we present the designed edge relations and multitask learning loss based on the VA theorem for multimodal emotion recognition, greatly improving the fusion efficiency on graphs.

\begin{figure*}[t]
\centering
\includegraphics[width=0.95\textwidth]{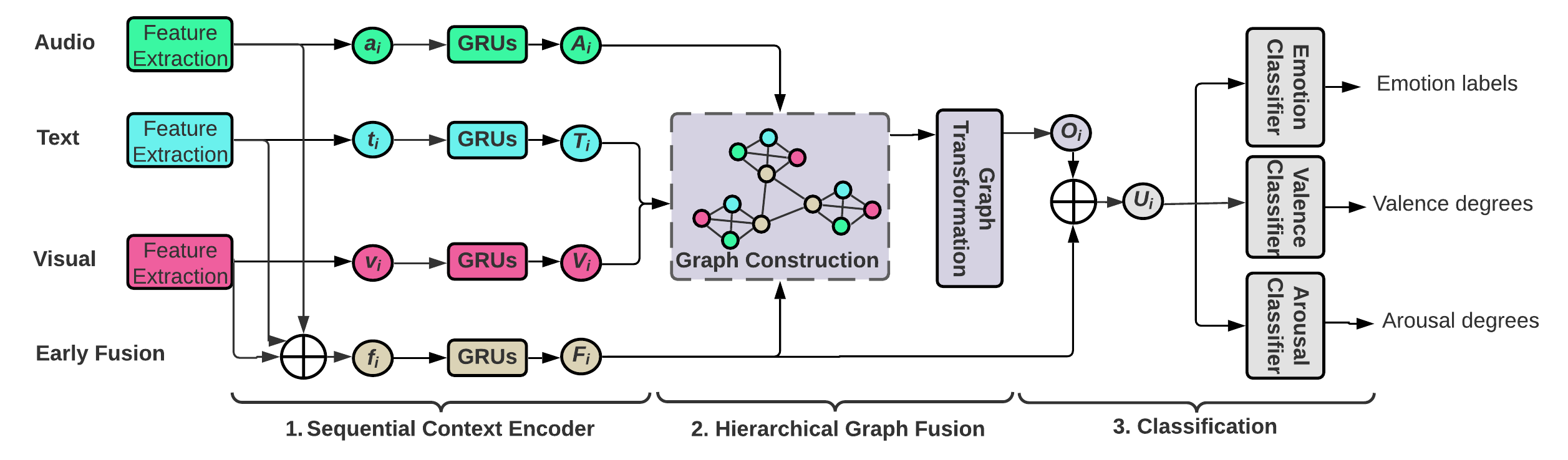}
\vspace{-10pt}
\caption{Framework of the proposed HFGCN method. It consists of a sequential context encoder, hierarchical graph fusion, and three classifiers.}
\vspace{-5pt}
\label{fig2}
\end{figure*}

\section{Feature Extraction} 
In this work, for feature extraction, following \cite{poria2017context}, audio, text, and video frame modalities are extracted from an input video by using individual pretrained networks transferred from other tasks. \textbf{Audio} features, such as the pitch and voice intensity of audio waveforms, are extracted using the OpenSMILE toolkit \cite{eyben2010opensmile} with IS13-ComParE \cite{schuller2013interspeech}.  \textbf{Text} features are embedded by Word2Vec vectors and then concatenated, padded, and further standardized to a one-dimensional vector by passing through a CNN \cite{karpathy2014large}. \textbf{Visual} features are extracted by 3D-CNNs \cite{ji20123d} pretrained from human action recognition to extract their body language. \textbf{Early fusion} features are obtained by concatenating the features extracted from audio, text, and video.

\section{Model Architecture}

In this section, we introduce the HFGCN model in detail. As shown in the framework of the proposed HFGCN in Fig.~\ref{fig2}, suppose there are $N$ utterances in a conversation, $i$ means the $i$-th utterance included in these utterances ($i \in 1,...,N$). We denote $a_i$, $t_i$, $v_i$, and $f_i$ as the extracted audio, text, visual, and their early fusion features, respectively. Their GRUs-encoded features are represented as $A_i$, $T_i$, $V_i$, and $F_i$, respectively. The graph transformed feature and final utterance representation are represented as $O_i$ and $U_i$, respectively. The task is to predict the emotion labels (\textit{happy, sad, neutral, angry, excited, frustrated, disgusted, and fear}) and VA degrees for each utterance by the classifiers. 

To achieve this goal, first, each modality's extracted feature and their early fusion feature $(a_i, t_i, v_i, f_i)$ are passed to the \textbf{sequential context encoder}. The output features $(A_i, T_i, V_i, F_i)$ are then passed to the \textbf{hierarchical graph fusion}, which conducts the graph construction and graph transformation. Finally, the graph transformed nodal feature $O_i$ is concatenated with the early fusion feature $F_i$ to obtain the final utterance representation $U_i$, which is then fed to the \textbf{classifiers}. Meanwhile, a multitask cross-entropy loss function is optimized to guide the emotion classification with the VA degrees.

\subsection{Sequential Context Encoder}
Similar to most proposed methods \cite{poria2017context, pan2020multi, majumder2018multimodal}, RNNs are used to capture the sequential contextual information that flows along the conversation. In the first part of our proposed model, the bidirectional gated recurrent units (GRUs) are used as the encoders to obtain the context-aware representations $A_i, T_i, V_i, F_i$: 
\begin{equation} \label{eq:1}
\begin{split}
% A_{i(+,-)1}
    A_i = \overleftrightarrow{GRUs}(a_i), 
    T_i = \overleftrightarrow{GRUs}(t_i), \\
    V_i = \overleftrightarrow{GRUs}(v_i), 
    F_i = \overleftrightarrow{GRUs}(f_i), \\
    \mathrm{for} \, i = 1,2,...,N,
\end{split}
\end{equation}
These features are then fed to the hierarchical graph fusion part for embedding the intermodality interactions and the relations between modality and emotion for the multimodal fusion processing.

\begin{figure*}[t]
\centering
\includegraphics[width=0.95\textwidth]{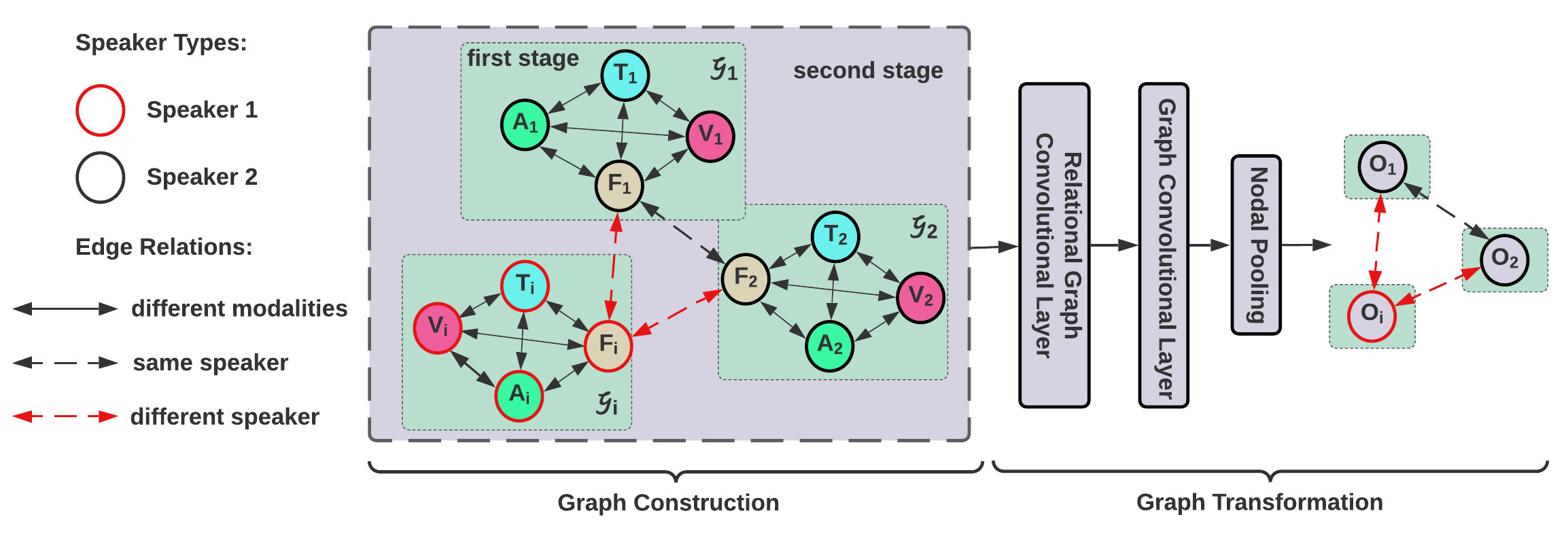} 
\vspace{-10pt}
\caption{Illustration of the Hierarchical Graph Fusion. The green boxes represent the proposed multimodal fusion that connects each modality with the solid arrows. In the utterance level (outside the gray boxes), each early fusion feature node $F_i$ is then connected with the dotted arrows. More detailed edge relation types are listed in Table.~\ref{tab1}. }
\label{fig3}
\end{figure*}

\subsection{Hierarchical Graph Fusion}

The hierarchical graph fusion processing is the key part of our proposed method. It consists of graph construction for multimodal fusion and graph transformation to accumulate neighbor nodes information. The details of the fusion method are shown in Fig.~\ref{fig3} and described as follows:

\subsubsection{Graph Construction:}
We use a two-stage construction to fuse the multimodal representations. As shown in Fig.~\ref{fig3}, $i$ denotes the $i$-th utterance of the $N$ utterances in the entire conversation. 
\textbf{In the first stage}, we denote $\mathcal{G}_i$ as a relational graph to encode the inter-modality interactions in the utterance. \textbf{In the second stage}, we expand the intermodality interactions to all the utterances by connecting all the first stage graphs $\mathcal{G}_1,\mathcal{G}_2,...,\mathcal{G}_N$ with their early fusion nodes $F_1,F_2,...,F_N$. We construct the graph $\mathcal{G}_i$ as follows:
\begin{equation} \label{eq:02}
\begin{split}
    \mathcal{G}_i = (\mathcal{N}_i, \mathcal{R}_{ij}, \mathcal{W}_{ij}), \\
    i, j = 1,2,...,N,
\end{split}
\end{equation}
where $\mathcal{N}_i$ represents the nodes consisting of the context-aware representations $A_i, T_i, V_i, F_i$ obtained from Eq.~\ref{eq:1}. $\mathcal{R}_{ij}, \mathcal{W}_{ij}$ are the edge relations and edge weights, which are designed as follows:

%\subsubsection{Edge Relations:}
\textit{\textbf{Edge Relations --}}
The edge relations $\mathcal{R}_{ij}$ are constructed depending on the connected nodes $ \mathcal{N}_i \in (A_i, T_i, V_i, F_i) $ and $ \mathcal{N}_j \in (A_j, T_j, V_j, F_j)$. The detailed relation types are listed in Table~\ref{tab1}. Relations 1--6 are designed for the first-stage graphs that are based on node modality, and relations 7--10 are designed for the second-stage graphs that are based on the utterances' positions and speakers. They together distinguish the contribution of each modality ($A_j, T_j, V_j$) to the emotion representation ($F_i$), which helps to capture the relations between modality and emotion. We summarize the edge relations based on the following three aspects:

 1). For modality dependency, within the same utterance ($i=j$), the relation depends on the modalities to which an edge is connected (i.e., node $A_1$ and node $V_1$ are connected with the edge relation 2, which indicates the interactions between audio and visual modality).

2). For speaker dependency, among different utterances ($i\not=j$) spoken by different speakers, the relation depends on the speaker to which an edge is connected (i.e., node $F_1$ that is uttered by speaker 1 and node $F_{3}$ that is uttered by speaker 2 are connected with the edge relation 9).

3). For temporal dependency, among different utterances ($i\not=j$) spoken by the same speaker, the relation depends on the relative position of different utterances (i.e., node $F_{1}$ and node $F_{2}$ are connected with the edge relation 7).

\begin{table}[t]
\centering
\begin{tabular}{l|l|l|l|l}
    \hline
    Relation & Speakers & Modality & Temporal & Example \\
    \hline
    1 & $s_1,s_1$ & A,T & i$=$j & $(A_1, T_1)$ \\
    2 & $s_1,s_1$ & A,V & i$=$j & $(A_1, V_1)$ \\
    3 & $s_1,s_1$ & A,F & i$=$j & $(A_1, F_1)$ \\
    4 & $s_1,s_1$ & T,V & i$=$j & $(T_1, V_1)$ \\
    5 & $s_1,s_1$ & T,F & i$=$j & $(T_1, F_1)$ \\
    6 & $s_1,s_1$ & V,F & i$=$j & $(V_1, F_1)$ \\
    7 & $s_1,s_1$ & F,F & i$<$j & $(F_1, F_2)$ \\
    8 & $s_1,s_1$ & F,F & i$>$j & $(F_2, F_1)$ \\
    9 & $s_1,s_2$ & F,F & i$<$j & $(F_1, F_3)$ \\
    10 & $s_1,s_2$ & F,F & i$>$j & $(F_3, F_1)$ \\
    
\end{tabular}
\caption{Detailed relation types between $i$-th utterance and $j$-th utterance with two distinct speakers and audio, text, visual, and early fusion nodes.} 
%($\bigcup_{i,j=1}^{N} r_{ij}$).}
\label{tab1}
\end{table}

%\subsubsection{Edge weights:}
\textit{\textbf{Edge weights --}} 
%\textbf{Edge eights}:
For the first-stage graph, within the same utterance ($i=j$), the edge weights $\mathcal{W}_{ij}$ between nodes $\mathcal{N}_i \in (A_i, T_i, V_i, F_i) $ and nodes $\mathcal{N}_j \in (A_j, T_j, V_j, F_j)$ are computed using a similarity-based attention mechanism as follows: 
\begin{equation} \label{eq:2}
\begin{split}
    \mathcal{W}_{ij} = \mathrm{softmax}(\mathcal{N}_i^TW_a\mathcal{N}_j),\\
    \mathrm{for} \,\, i, j = 1,2,...,N,
    % \mathrm{for} \, i, j \in [k,k+1,k+2,k+3], k = 1,...,N, %b = \mathcal{N}_i-a.
\end{split}
\end{equation}
where $W_a$ denotes the learning weights of the attention mechanisms. The $\mathrm{softmax}$ function ensures that for the node of each modality, the edges connected with the remaining modalities have a total attention weight of $1$. For example, audio modality node $A_i$ receives a total contribution of 1 from $T_j, V_j, and F_j$.

For the second-stage graph, the utterances are connected to each other with early fusion nodes $F_i$. 
The second stage graph's edge weights $\mathcal{W}_{ij}$ between node $F_i$ and node $F_j$ are then calculated based on a multilayer perceptron (MLP) \cite{wu2019simplifying} as follows:
\begin{equation} \label{eq:3}
\begin{split}
    \hat{F_i}, \hat{F_j} = \mathrm{Linear}(F_i), \mathrm{Linear}(F_j),\\
    \mathcal{W}_{ij} = \mathrm{softmax}(\mathrm{tanh}(\mathrm{concat}(\hat{F_i}, \hat{F_j}) W_b),\\
    % \mathrm{for} \, x = F_i, y = F_j, \\
    \mathrm{for} \,\, i, j = 1,2,...,N,
\end{split}
\end{equation}
$F_i$ and $F_j$ are first passed to two linear layers to obtain the query $\hat{F_i}$ and key $\hat{F_j}$. $W_b$ denotes the learning weights of the attention mechanisms. Similarly, the softmax function ensures that all the edges within the utterance window have a total attention weight of 1. This can be understood as a general multiplicative application of the attention of neighboring nodes.

\subsubsection{Graph Transformation:}
%\textbf{Graph Transformation}:
As shown in Fig.~\ref{fig3}, after graph construction, the graph $\mathcal{G}$ is fed to the graph transformation part, which consists of a relational graph convolutional layer, a graph convolutional layer, and a nodal pooling layer.

The first relational graph convolutional layer \cite{sukhbaatar2015end} encodes the modality dependencies and speaker dependencies and outputs the transformed graph features $O^{(1)}_i$ using the edge relations $\mathcal{R}_{ij}$ with the input nodes $\mathcal{N}_i \in (A_i, T_i, V_i, F_i)$ and $\mathcal{N}_j \in (A_j, T_j, V_j, F_j)$. The second graph convolutional layer uses $O^{(1)}_i$ as input and further accumulates the local neighborhood information and outputs the transformed graph features $O^{(2)}_i$ as follows:
\begin{equation} \label{eq:5}
\begin{split}
    O_{i}^{(2)} = \mathrm{ReLU}(\sum_{j\in \mathcal{N}^r_i}W^{(2)}_cO_j^{(1)}+W_d^{(2)}O_i^{(1)}), \\
    O_{i}^{(1)} = \mathrm{ReLU}(\sum_{r\in \mathcal{R}_{ij}}\sum_{j\in \mathcal{N}^r_i}\frac{\mathcal{W}_{ij}}{|\mathcal{N}^r_i|}W^{(1)}_c \mathcal{N}_j+\mathcal{W}_{ii} W_d^{(1)} \mathcal{N}_i), \\
    \mathrm{for} \,\, i = 1,2,...,N,
\end{split}
\end{equation}
where $\mathcal{W}_{ij},\mathcal{W}_{ii}$ are the edge weights of the first transformation, and $W^{(1)}_c, W^{(1)}_d, W^{(2)}_c, W^{(2)}_d$ are the independent learning weights of the first and second transformations. $\mathcal{N}^r_i$ denotes the neighbor indices of vertex $j$ under the relation $r \in \mathcal{R}_{ij}$. $|\mathcal{N}^r_i|$ is set as the normalization constant.

This two-step graph transformation effectively accumulates the normalized sum of the neighborhood information that is enriched with modality and speaker dependencies for each utterance. Because each utterance has four nodes, nodal pooling is performed to reduce the total number of nodes from $4N$ to $N$. For each first stage graph that has 4 nodes ($\mathcal{N}_j=  A_j, T_j, V_j, F_j$), it will output 1 node after the pooling. We applied global average pooling to obtain the $O_i$.
\begin{equation} \label{eq:4.5}
\begin{split}
    O_{i} =  \mathrm{AvgPooling}(O_{i}^{(2)}),\\%,O_{i+1}^{(2)},O_{i+2}^{(2)},O_{i+3}^{(2)}])\\
    \mathrm{for} \, i = 1,2,...,N,
\end{split}
\end{equation}

\subsection{Classification}
For the classification part, we obtain the final utterance representation $U_i$ for the $i$-th utterance as follows: 
\begin{equation} \label{eq:6}
\begin{split}
    U_i = \mathrm{concat}[F_i, O_i],\\
\end{split}
\end{equation}
where $O_i$ is the graph transformed feature obtained in the Graph Transformation section, and $F_i$ represents the GRUs-encoded early fusion feature processed in the Feature Extraction section. Finally, all the final utterances' representations $U_1,...,U_N$ are then passed to three fully connected networks: emotion classifier, valence classifier, and arousal classifier. Each includes a fully connected layer with a softmax function to output the probability distribution of the emotion labels, valence degrees, and arousal degrees. 

\subsection{Multitask Training Setup}
Based on Psychology, the VA model \cite{Barrett1998-BARDEO,barrett2006solving,article} suggests that the emotions can be categorized or described based on a two-dimensional VA subspace. Thus, we design a special loss function $L_{e,v,a}$ for jointly predicting the emotion labels and VA degrees, which is the categorical cross-entropy $\mathcal{L}$ for emotion classification and VA classification as follows:
\begin{equation} \label{eq:7}
\begin{split}
    L_{e,v,a} = \mathcal{L}(\mathcal{P}_{i,j}^{e}, y_{i,j}^e)+W_1\mathcal{L}(\mathcal{P}_{i,j}^v, y_{i,j}^v)+W_2\mathcal{L}(\mathcal{P}_{i,j}^a, y_{i,j}^a),\\
    \mathcal{L}(\mathcal{P}_{i,j}, y_{i,j}) = -\frac{1}{\sum^{C}_{i=1}N(i)}\sum^{C}_{i=1}\sum^{N(i)}_{j=1}log\mathcal{P}_{i,j}[y_{i,j}], 
\end{split}
\end{equation}
where $C$ is the number of conversations/dialogues, $N(i)$ is the number of utterances in the $i$-th conversation, $\mathcal{P}^{e}_{i,j},\mathcal{P}^{v}_{i,j},\mathcal{P}^{a}_{i,j}$ are the model output probability distributions of the emotion labels, valence degrees, and arousal degrees of the $j$-th utterance in the $i$-th conversation, $y^{e}_{i,j},y^{v}_{i,j},y^{a}_{i,j}$ are their corresponding expected labels, and $W_1, W_2$ are the loss terms' weights that are set as hyperparameters.

As the VA degrees are discrete but ordinal, we experimented on our multitask loss with both regression type loss and classification loss. The cross entropy was found to be significantly better than other types of losses.

\section{Experiments}

\subsection{Datasets}

\begin{table}
    \caption{Distribution of training and test sets. We used 10\% of the training dialogues as the validation set. For IEMOCAP, we followed the previous work \cite{pan2020multi} to use the first four sessions as the training set and the last session as the test set.}
    %  \vspace{-0.2cm}
    \begin{tabular}{c c c c} 
    %   \toprule
        \hline
        Dataset  & Split & Utterance & Dialogue \\ 
    %   \midrule
        \hline
        \hline
       IEMOCAP & Train/Val & 5810 & 120\\ 
               & Test & 1623 & 31 \\
        \hline
       MELD & Train/Val & 11098 & 1153 \\ 
            & Test & 2610 & 280  \\ 
       \hline
    %   \bottomrule
    \end{tabular}
    \centering
    \label{tb:dataset}
    % \vspace{-0.5cm}
\end{table}

We evaluated our HFGCN model on two benchmark datasets: IEMOCAP \cite{busso2008iemocap} and MELD \cite{poria-etal-2019-meld}. Both datasets are multimodal datasets containing text transcription, audio waveform, and visual frame information for every utterance of each conversation/dialogue. They are split into training, test, and validation sets, as shown in Table~\ref{tb:dataset}.

The \textbf{IEMOCAP} \cite{busso2008iemocap} dataset contains 10K videos split into 5 min of dyadic conversations for human emotion analysis. Two speakers participated in the conversation. Each conversation is split into spoken utterances. Each utterance in every dialogue is annotated using an emotion label. To align with previous works, we implemented both six-class and four-class AER on IEMOCAP, which are \textit{angry, happy, excited, sad, frustrated, and neutral} and \textit{angry, happy(excited), sad(frustrated), and neutral}.  In addition, each utterance was also annotated by two evaluators for the valence and arousal degrees, which range from one to five. We took their average to obtain the final VA degrees.

The \textbf{MELD} \cite{poria-etal-2019-meld} dataset contains more than 1.4K dialogues and 13000 utterances from the Friends TV series. Multiple speakers participated in the conversations. Each utterance in every dialogue is annotated as one of the seven emotion classes: \textit{anger, disgust, sadness, joy, surprise, fear, or neutral}. In contrast to IEMOCAP, we used MELD for evaluating only the emotion labels, and not the VA degrees. Thus, the HFGCN-VA that includes the multitask training was not performed on the MELD. 

\begin{table*}
	\caption{BC-LSTM, DialogueRNN, and DialogueGCN are designed for the unimodal AER. Thus, we reimplemented them in a multimodal setting by simply concatenating the features of the three modalities. Other models left with \textbf{-} in the table are non-open-sourced. Therefore, we can only report the published results from their respective papers.}
	\centering
	\vspace{-5pt}
	\begin{tabular}{c |c c c c c c | c || c c || c}
		\hline
		\multicolumn{1}{c|}{Methods} & \multicolumn{7}{c||}{IEMOCAP (6 class)} & \multicolumn{2}{c||}{IEMOCAP (4 class)} & \multicolumn{1}{c}{MELD} \\
		\cline{2-10}
		\multicolumn{1}{c|}{} & \multicolumn{1}{c}{happy} & sad & neutral & angry & excited & frustrated & avg. F1 & avg. F1 & avg. acc & avg. F1 \\
		\hline
		BC-LSTM &  34.43  & 60.87  & 51.83  & 56.73  & 57.95 & 58.92 & 54.95 & 60.41 & 61.50  & 56.80\\
		%ICON & 29.91 & 64.57 & 57.38 & 63.04 & 63.42 & 60.81 & 58.54 & 65.40 & 66.01 & -\\
		DialogueRNN & 39.16 & \textbf{81.69} & 59.77 & 67.36 & 72.91 & 60.27 & 64.58 & 70.85 & 70.68 & 57.11 \\
		DialogueGCN & 47.1  & 80.88  & 58.71  & 66.08  &  70.97 & 61.21 & 65.04 & 71.33 & 71.29 & 58.23  \\
		MMAN &  -  & -  & -  & -  & - & - & - & 74.11 & 73.94 & -\\
		MMGCN & 42.34  & 78.67  & 61.73 & \textbf{69.00} & 74.33 & \textbf{62.32} & 66.22 & - & - & 58.65 \\
		\hline
% 		[0.60645161 0.79847909 0.63276836 0.63384615 0.75389408 0.60952381]
		HFGCN & \textbf{60.65}  & 79.85  & \textbf{63.28}  & 63.38 & \textbf{75.38} & 60.95 & \textbf{67.24} & \textbf{74.90} &\textbf{74.68} & \textbf{59.71}\\
% 		[0.54736842 0.82466281 0.67556742 0.63687151 0.76069731 0.60511364]
		HFGCN-VA & 54.74  & 82.47 & 67.56 & 63.69 & 76.07 & 60.51 & 68.18 & 75.98 & 75.91 & - \\
		\hline
	\end{tabular}
	\vspace{-5pt}
	\label{result}
\end{table*}

\begin{figure*}[t]
\centering
\includegraphics[width=1.0\textwidth]{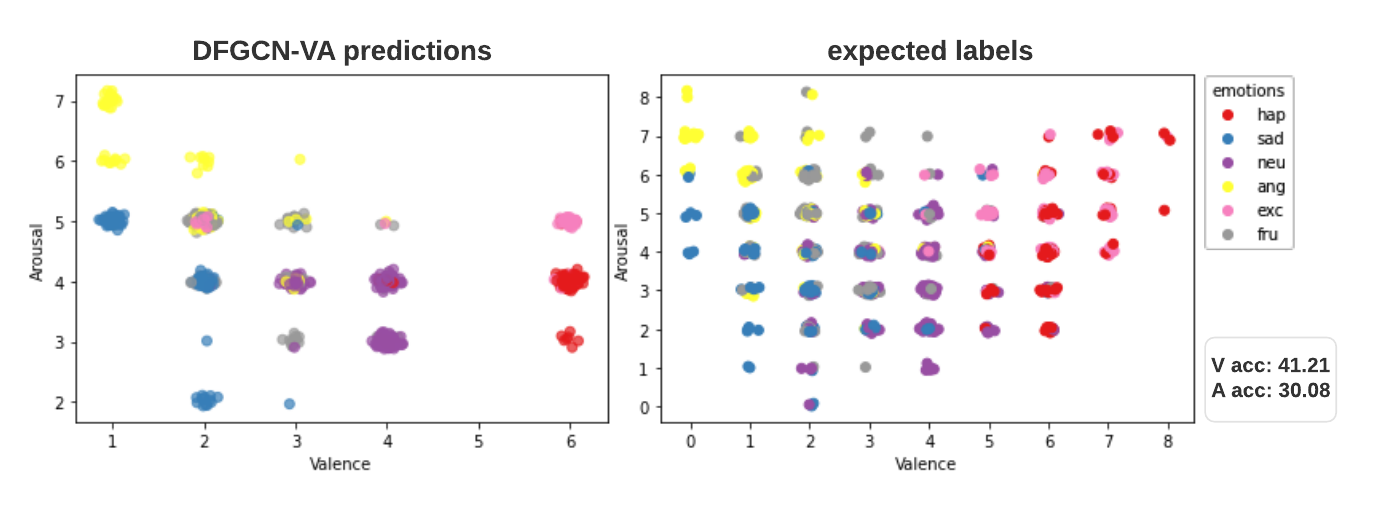} % Reduce the figure size so that it is slightly narrower than the column.
\vspace{-25pt}
\caption{Emotion states projected on the VA subspace for the 1000 utterances randomly sampled from the IEMOCAP test dataset. Small random Gaussian noise is added to avoid over-clustering. The valence and arousal prediction accuracy scores are reported in the bottom-right corner.}
\vspace{-10pt}
\label{fig4}
\end{figure*}

\subsection{Baseline systems}
To evaluate the proposed method, we compared the results with the following models for AER tasks.  
For the open-sourced models, we reimplemented them if they did not report some results on the datasets. 

\textbf{BC-LSTM}: \cite{poria2017context} 
This is the baseline method, in which context-aware utterance representations are generated by capturing the contextual content from the surrounding utterances using a Bi-directional LSTM network.

\textbf{DialogueRNN}: \cite{Majumder2019DialogueRNNAA}
This is a recurrent network that uses two GRUs to track individual speaker states and context for emotion recognition in conversation.

\textbf{MMAN}: \cite{pan2020multi} This is a state-of-the-art LSTM-based multimodal AER method that consists of a hybrid fusion LSTM network with multimodal attention mechanisms.

\textbf{DialogueGCN}: \cite{ghosal2019dialoguegcn} This is a baseline graph convolutional network that focuses on the textual modality for emotion recognition. 

\textbf{MMGCN}: \cite{hu2021mmgcn} This is a recent state-of-the-art multimodal AER method that uses GCN to fuse multimodal information with the intermodality interactions. 

\textbf{HFGCN variants}: This is our proposed method that is efficient with intermodality interactions. HFGCN-VA is the proposed HFGCN with the VA classifiers and is trained under the multitask training setup. For a fair comparison, we did not compare the HFGCN-VA that uses additional valence-arousal degrees as extra training data, but only used it for interpretation.

\subsection{Implementation Details}
We implemented our proposed HFGCN model on the PyTorch framework \footnote{https://pytorch.org/} and PyTorch geometric for the Hierarchical Graph Fusion parts. During training, we chose the Adam optimizer \cite{kingma2014adam} and set the hyperparameters as follows: learning rate of 0.0001, 50 epochs with early stopping, dropout rate of 0.35, loss term weights $W_1 = 0.15, W_2 = 0.15$.

\subsection{Comparison and Analysis}
The experimental results for the proposed HFGCN, HFGCN-VA, and previous approaches are shown in Table~\ref{result}. To align with previous studies, we report the F1 scores for each emotion class and evaluate the overall classification performance using their weighted averages of the IEMOCAP with six classes. For the IEMOCAP with four classes, we report both the average accuracy scores and weighted F1 scores. For the MELD, we report the weighted F1 scores. 

In Table~\ref{result}, our HFGCN performs better than the compared methods. HFGCN attains the best overall performance with approximately +1\% among all the average scores compared to the previous state-of-the-art methods MMGCN and MMAN. In particular, our HFGCN attains a large improvement on \textit{happy}, which is around 13\% compared to the previous best model in classifying happy emotion. We note that happy is the emotion with the least utterances, which demonstrates the ability of HCGCN to recognize minority emotion classes. 

Compared to the RNN-based methods, our model significantly improves the results among all datasets, which is possibly due to most of the RNNs' being ignorant of the \textbf{intermodality interactions}. However, for the \textit{sad}, which is the emotion with the most utterances, the DialogueRNN holds a better F1 score. By checking their confusion matrix, we notice that the DialogueRNN misclassifies a large proportion of other minority emotion classes to the sad emotion.

Compared to the GCN-based methods, our HFGCN also improves the classification results of both average and minority emotion classes (happy, neutral, excited). HFGCN utilizes the hierarchical fusion that consists of early fusion and graph fusion, which outperforms the MMGCN that only uses graph fusion and DialogueGCN that only utilizes early fusion.

We observed that the additional VA information does improve the results by about 1\% among the average F1 scores and accuracy scores. Through the multitask learning, the HFGCN-VA provides the emotional states' visualization and provides better error analysis and interpretation.

\subsection{Ablation Study}

We conducted the ablation study for different stage graphs in Table ~\ref{ablation1}. 
Removing both of the graphs results in a drastic decrease in performance. It is equivalent to removing the entire GCN part, which is similar to the BC-LSTM \cite{poria2017context}.
For the removal of the first-stage graphs, we only include the early fusion nodes $F$ during the graph construction, which is similar to the DialogueGCN \cite{ghosal2019dialoguegcn} with early fused inputs. It fails to capture the \textbf{intermodality interactions} as the early fusion nodes do not capture the correlations between each modality. This proves our graph fusion method's superiority over the traditional early fusion method. For the removal of the second-stage graphs, the model is similar to the MMGCN \cite{hu2021mmgcn} that, during the graph construction, the unimodality nodes connect to one other and their future/past utterances' of unimodality nodes. 
Without the presence of the early fusion nodes, it fails to capture the \textbf{relations between modality and emotion} as the early fusion nodes connect the first stage graphs and distinguish their contribution to the entire utterance representations. This proves that hierarchically combining graph fusion and early fusion (HFGCN) is more effective than using graph fusion (MMGCN) only.
 
Further, we present the effect of edge relations and edge weights in Table ~\ref{ablation2}. Removing both of the attention-based edge weights and edge relations is equivalent to a vanilla GCN with undirected, unweighted edges, which results in around a 5\% drop in the F1 scores. For the removal of the attention-based edge weights, the result drops about 2.7\%. For the removal of the edge relations, the results drops about 3.9\%, which indicates that the edge relations have more overall importance. 

\begin{table}[!htp]
    \caption{Ablation results w.r.t the construction of first stage graph and second stage graph on the IEMOCAP dataset (4 class).}
    %  \vspace{-0.2cm}
    \begin{tabular}{c c c} 
        \hline
        1st stage graph & 2nd stage graph & avg. F1\\ 
        \hline
        \xmark & \xmark & 62.07\\ 
        \checkmark & \xmark & 72.95\\
        \xmark & \checkmark & 71.57\\
        \checkmark & \checkmark & 74.90\\
        \hline
    \end{tabular}
    \centering
    \label{ablation1}
\end{table}

\begin{table}[!htp]
    \caption{Ablation results w.r.t the attention-based edge weights and edge relations on IEMOCAP dataset (4 class).}
    %  \vspace{-0.2cm}
    \begin{tabular}{c c c} 
        \hline
        edge weights attn & edge relation & avg. F1\\ 
        \hline
        \xmark & \xmark & 70.10\\ 
        \checkmark & \xmark & 71.04\\
        \xmark & \checkmark & 72.15\\
        \checkmark & \checkmark & 74.90\\
        % \xmark & \xmark & \xmark & 62.07\\ 
        % \checkmark & \xmark & \xmark & 72.95\\
        % \xmark & \checkmark & \xmark & 71.57\\
        % \checkmark & \checkmark & \xmark & 74.90\\
        % \checkmark & \checkmark & \checkmark & 74.90\\
        \hline
    \end{tabular}
    \centering
    \label{ablation2}
\end{table}

\subsection{Visualization of Emotional States}
As shown in Fig. ~\ref{fig4}, we demonstrate our VA multitask training's visualization, which is corresponded with the original VA model \cite{Barrett1998-BARDEO}. For those emotions with low valence degrees, they are usually sad or frustrated with low arousal degrees while angry with high arousal degrees. For those emotions with high valence degrees, they are usually happy with low arousal degrees while excited with high arousal degrees. The neutral emotion usually has intermediate VA degrees. 

Let us carry out the error analysis. Compared to the expected labels, most of the predictions are over-generalized. For example, the happy and excited emotions are only in large clusters at valence degree = 6 while, in fact, they are much more diverse in the arousal axis. Furthermore, most of the predicted frustrated emotions tend to have lower arousal degrees than the actual ones. At valence degree = 6, arousal degree = 5, our model misclassifies many excited emotions as angry and frustrated emotions, which is unusual as the excited emotions have large valence degrees. After manually checking those misclassified excited results, we notice that most of them are short utterances such as ``Okay'' or ``I just can't'' from the angry emotions, which lie on the transition between neutral and angry. The emotion visualization shows a general idea about how the emotions are distributed on the valence-arousal subspace and suggests that we need to pay more attention to the over-generalized arousal axis.

\section{Conclusions}
In this paper, we proposed HFGCN, a hierarchical fusion graph convolutional network for better multimodal emotion recognition. HFGCN is equipped with a novel graph fusion method that consists of two-stage graph construction, attention-based edge weights, and relational graph transformation that captures the intermodality interactions. We also presented a multitask loss to guide the joint prediction of emotion labels and valence-arousal (VA) degrees. Extensive experiments on two public emotion recognition datasets show the effectiveness of our approach. Ablation studies and visualizations further demonstrate the efficacy of each component of our HFGCN model. Additionally, we provide insightful analysis and interpretation by projecting the emotional states to a 2D VA space. While the model is designed for AER, we believe that our method can be generalized to other classification tasks that need to fuse information from various multiple modalities. 

\bibliography{aaai22.bib}

\end{document}